\documentclass[aps,showpacs]{revtex4}

\usepackage{amsmath}
\usepackage{graphicx}

\newcommand{\sub}[1]{{\mbox{\scriptsize #1}}} 

\begin{document}

\preprint{Single-atom scattering in waveguides}

\title[Single-atom scattering in waveguides]{Quantum description of
  light pulse scattering on a single atom in waveguides}

\author{Peter Domokos}
\altaffiliation{On leave from Research Institute for Solid State Physics and
   Optics, Budapest, Hungary}
 \email{Peter.Domokos@uibk.ac.at}

\author{Peter Horak}
\altaffiliation{Present address: Optoelectronics Research Centre, University of
Southampton, United Kingdom}

\author{Helmut Ritsch}%
\affiliation{%
Institut for Theoretical Physics, Universit\"at Innsbruck,
Technikerstr.~25, A-6020 Innsbruck,\\ Austria}%

\begin{abstract}
We present a time dependent quantum calculation of the scattering of a
few-photon pulse on a single atom.  The photon wave packet is assumed to
propagate in a transversely strongly confined geometry, which ensures strong
atom-light coupling and allows a quasi 1D treatment. The amplitude and phase of
the transmitted, reflected and transversely scattered part of the wave packet
strongly depend on the pulse length (bandwidth) and energy. For a transverse
mode size of the order of $\lambda^2$, we find nonlinear behavior for a few
photons already, or even for a single photon. In a second step we study the
collision of two such wave packets at the atomic site and find striking
differences between Fock state and coherent state wave packets of the same
photon number.
\end{abstract}


\maketitle

\section{Introduction}

Identifying and realizing systems with strong coupling between light
and matter is one of the central objectives of current research in
quantum optics.  In the strong coupling regime the coherent
interaction between a few atoms and the radiation field of few photons
takes place on a fast time scale and is not masked by the dissipative
coupling to the environment. A weak quantum field can change the
internal quantum state of an atom and, vice versa, a single atom
is able to have an appreciable influence on the light field. Both
effects have, evidently, many interesting practical implications
including single atom detectors as well as nonlinear micro-optical
elements on the single-photon level, which are in the heart of
all-optical quantum information processing schemes \cite{Torma},
e.g.~to facilitate Bell state detectors for photons \cite{Kwiat}.

So far optical experiments in the strong coupling regime have been
performed with atoms put in tiny high-Q cavities
\cite{pinkse00,hood00}. Here the origin of the strong coupling is the
small volume of the discrete cavity modes. The local electric field
sustained by a single photon becomes then very large. Strong coupling
in the microwave regime yielded several spectacular results
\cite{Haroche,Walther}.  However, the electric field enclosed in three
dimensions allows only limited accessibility, which complicates direct
detection of the light and limits the time scales.  In addition, the
scalability to a large number of such elements seems rather difficult.

Hence, there is still a high demand for alternative setups which allow
controlled manipulation and strong (nonlinear) coupling of single
photons. It seems natural to try to lift the confinement of the light
field in one particular dimension and consider the interaction of
transversely confined propagating fields with nonlinear optical
elements. If a photon forms a very short wave packet along its
propagating direction, we again get a small total volume and a high
field per photon. Conceptually the most simple nonlinear element for
such wave packets is of course a single atom with a resonant dipole
transition. An atomic dipole can be associated with an effective
radiative cross section of $\sigma_A=3 \lambda^2/2\pi$.  Hence, as a
first guess one would expect that it is sufficient to simply focus a
pulsed light beam down to a cross section of the same order of
magnitude as $\sigma_A$ to enter the strong coupling regime.  This is
not too restrictive as beams can be focused down to a spot size
smaller than $\lambda^2$. Unfortunately, this line of argumentation is
too naive and a precise calculation for such a strongly focused beam
shows that the light will propagate through the atom without being
noticeably influenced \cite{vanEnk}. The reason for this can be traced
back to the large range of transverse wave vectors involved in the
dynamics yielding rapid variation of input phase and polarization.
Another way to enhanced effective atomic cross section could, in
principle, consist in using a quantum degenerate gas as an optically
dense medium \cite{lukin}. Here the use of many atoms leads to quite
stringent practical limitations.

It is the advent of highly developed nanofabrication techniques that
may open the way to new geometries avoiding these problems and
facilitating setups to implement quasi one-dimensional scattering
processes on dipoles. Miniaturized waveguides on surfaces or optical
fibers can be fabricated with cross sections on the order of the
optical wavelength square.  Photons traveling within are well confined
in the transverse direction, while the field remains uniform in the
longitudinal direction of propagation. The problems of the beam
divergence and rapidly varying wave vector, encountered in free space,
are missing in the waveguide setup. Hence one can hope that a single
atom placed in the field of such a mode will have considerable
influence on the dynamics. The investigation of the prospects of such
a setup is the subject of the present paper.

We will study the time dependent interaction of an atomic point dipole
modeled by a two-level atom and the quantized electromagnetic field in
a single transverse mode waveguide.  We use the term ``waveguide'' in
a general sense, avoiding any precise geometric specification.
Instead, we set up a model that accounts for the generic features of
these devices. The central property is that the atom interacts with a
continuum of transversally confined propagating modes. A similar
system has been studied to investigate the influence of input
photon-statistics on single-atom absorption \cite{kochan}. For this
the theory of cascaded open systems \cite{gardiner,carmichael} as an
extended form of the input-output formalism was applied. In contrast
to the stationary scattering scenario, we will concentrate here on the
time-dependence of the scattering process, i.e. we consider {\it wave
  packets} consisting of one or a few photons impinging on an atom.
This allows to find the dependence of the atomic dynamics and the
amount of scattering on the light pulse bandwidth and energy beyond
the narrow-bandwidth limit considered in Ref.~\cite{kochan}.  For very
short pulses the energy of a single photon is strongly concentrated in
space and thus saturation is expected to prominently influence the
scattering process even for weak energy incident wave packets. As a
consequence, we have to take the full quantum dynamics of the dipole
into account, including saturation to all orders.

To treat this problem of wave packet scattering we develop a method
based on a set of Heisenberg-Langevin equations for the field and
atomic operators.  These equations, relying on first principles of
quantum mechanics, will be transformed into a simplified form by means
of a Markoff approximation. This approximation closely relates our
approach to the cascaded open systems method.  Nevertheless we are
primarily interested in the field dynamics, called ``channel'' in the
scheme of cascaded systems. Hence we will keep the waveguide electric
field as an explicit dynamic component of our model. Let us mention
here that some numerical examples calculating the electric field of
the propagation of a single-photon wave packet through a single atom
have been obtained before \cite{drobny}.  Similarly the dynamics of a
2D 1-photon wave packet and many two-state atoms modeling a beam
splitter was numerically solved \cite{havukainen}.

Nevertheless, the systematic and exhaustive discussion of the
parameter space by a purely numerical approach seems hopeless.
Fortunately in our method, we are able to get explicit analytical
expressions for the scattered waveguide field. This result exhibits
the scaling with the ratio of the transverse mode size and the atomic
cross section. Thus we can easily reveal how this ratio governs the
energy redistribution in a scattering process. In addition, physical
insight in the phase properties and the pulse shape deformation can
also be gained.

As a possible application of our model we address the problem of
atom-mediated nonlinear photon-photon interaction. So far, significant
coupling has been reached with the help of a high-Q cavity
\cite{lange} or might be reached by increasing the number of atoms
\cite{lukin,chiao}.  The waveguide-atom interaction could be the basis
of nonlinear pulse amplifiers or even that of photonic quantum gates.
Here we limit the study to a prototype interference experiment in
which two wave packets impinge simultaneously on the same atom.

The paper is organized as follows. In Section II we present the model
and introduce the basic parameter describing the coupling of the
atomic dipole and the waveguide, and finally, calculate the scattered
electric field generally within a Markoff approximation. In Section
III the transmitted and reflected Poynting vectors are studied for
various initial states of the wave packet. These states include
coherent states and the single-photon Fock state.  We study the role
of the pulse bandwidth in the scattering process and the saturation
nonlinearity for multi-photon wave packets. The phase properties and
the pulse shape deformation is also discussed in this section. Then,
Section IV is devoted to the interaction of light pulses. We conclude
finally in Section V.

\section{The model}

Let us consider a single branch of modes propagating in the $+z$
direction of a lossless waveguide. The longitudinal wave number $k$ of
the modes is assumed to obey the simple dispersion relation
$k=\omega/c$, which is valid far from the branch threshold
\cite{snyder}. The effective transverse cross section of the modes
$$
{\cal A} = \frac{\int dx\,dy\,|f_k(x,y)|^2}{|f_k(x_a,y_a)|^2},
$$
where $f_k(x,y)$ is the transverse mode function and $(x_a,y_a)$
the position of the atom, is approximately constant in the range of
the relevant longitudinal wave numbers. The details of the transverse
mode profile do not play an important role in the forthcoming
calculations as long as the atom may be treated as point-like. 
Only the value of the mode function at the position of the atom enters
in the calculation. This value has been incorporated in the definition
of the effective cross section ${\cal A}$. Similarly, the polarization
properties of the field can also be omitted provided the field is
closely uniform across the spatial wave function of the atom. No
assumption is made on the atomic position relative to the waveguide.
In principle the atom can sit inside a hollow waveguide, it can be
embedded in the dielectric material of a fiber or can be even outside
a dielectric interacting with the evanescent field close to the
surface.

The continuum field quantization follows the theory presented in
Ref.~\cite{Blow}. The electric radiation field is decomposed into
positive and negative frequency parts
\begin{equation}
  \label{eq:E}
  E(z,t) = E^{(+)}(z,t) + E^{(-)}(z,t) \; ,
\end{equation}
where
\begin{equation}
  \label{eq:E+}
  E^{(+)}(z,t)= i \int_0^{\infty} d\omega \sqrt{\frac{\hbar
      \omega}{4 \pi \epsilon c {\cal A}}} \left( a_\omega(t)
    e^{-i\omega(t-z/c)} + b_\omega(t) e^{-i\omega(t+z/c)} \right) 
  \; .
\end{equation}
We separated the modes of the two counter-propagating directions into
sets of $a_\omega$ and $b_\omega$, the first corresponding
to the forward, the latter to the backward propagating modes. The
electric field is given in the interaction picture. The field
amplitude variables $a_\omega(t)$ and $b_\omega(t)$ describe
then the time variation due to the interaction with the atom. They
follow the usual bosonic commutation rules,
\begin{equation}
  \label{eq:commutation}
  \left[a_\omega(t), a_{\omega'}^\dagger(t)\right] = 
  \left[b_\omega(t), b_{\omega'}^\dagger(t)\right] =
  \delta(\omega-\omega')\; ,
\end{equation}
all the other commutators vanish.

The atomic dipole, again in interaction picture, reads 
\begin{equation}
  \label{eq:d}
  d = d_{eg} (\sigma_- e^{-i\omega_A t}+\sigma_+
  e^{i\omega_A t}) \; ,
\end{equation}
where the operators $\sigma_\pm$ together with
$\sigma_z=(\sigma_+\sigma_- -\sigma_-\sigma_+)/2$ form a pseudospin
obeying the spin-$\frac{1}{2}$ algebra. We assume that the dipole
moment is oriented parallel to the field polarization at the atomic position
yielding maximum coupling.

The dipole interaction Hamiltonian in the rotating wave approximation
is given by
\begin{equation}
  \label{eq:H_int}
  H_I = -i \hbar \int d\omega g_\omega \left( \sigma_+ 
    (a_\omega e^{i\omega z_A/c} + b_\omega e^{-i\omega z_A/c}) 
    e^{-i(\omega-\omega_A) t } - h.c \right)\,,
\end{equation}
where $z_A$ is the position of the atom, and the coupling constants are
\begin{equation}
  \label{eq:g}
  g_\omega = \sqrt{\frac{\omega}{4 \pi \epsilon \hbar c {\cal A}}} d_{eg}\,.
\end{equation}
Note that the dimension of the coupling constant $g_\omega$ is not a
frequency but $1/\sqrt{\mbox{sec}}$. The use of the rotating wave
approximation is justified as we consider light pulses with bandwidth
much smaller than the frequency $\omega_A$.

The evolution of the system variables is governed by a set of coupled
Heisenberg-Langevin equations \cite{cct}
\begin{subequations}
  \label{eq:HL}
\begin{align}
  \frac{d}{dt} a_\omega &= g_\omega \hat\sigma_- e^{-i\omega
    z_A/c} e^{i(\omega-\omega_A) t} \,,\\
  \frac{d}{dt} b_\omega &= g_\omega \hat\sigma_- e^{i\omega
    z_A/c} e^{i(\omega-\omega_A) t} \,,\\
  \frac{d}{dt} \sigma_- &= -\gamma_0 \sigma_- +
    2\sigma_z \int d\omega g_\omega (a_\omega e^{i\omega z_A/c} + b_\omega e^{-i\omega z_A/c}) 
    e^{-i(\omega-\omega_A) t} + \hat\xi_- \,,\\
  \frac{d}{dt} \sigma_z &= -2\gamma_0 (\hat\sigma_z+1/2)-
    \int d\omega g_\omega \left(\sigma_+ 
    (a_\omega e^{i\omega z_A/c} + b_\omega e^{-i\omega z_A/c}) 
    e^{-i(\omega-\omega_A) t } + h.c \right) + \hat\xi_z \; .
\end{align}
\end{subequations}
Besides the terms originating from the interaction Hamiltonian
(\ref{eq:H_int}), we account for the interaction of the atom with an
environment. This coupling results in a dissipation process with decay
rate $\gamma_0$ and with associated noise represented by the operators
$\hat\xi$. The physical role of this $\gamma_0$ decay process is that it
provides a channel for the transverse scattering, i.e., when photons
are scattered by the atom out of the waveguide. The environment
consists of the free space radiation modes \footnote{For an atom
  within the waveguide, the lossy modes, not guided by the structure,
  compose the environment. These modes are continued into the free
  space radiation modes outside the material. For small numerical
  aperture the mode density of the lossy modes is close to the one in
  free space}. 
Of course, the presence of the 1D waveguide slightly
perturbs the surrounding mode structure with respect to the simple
free space case. This effect, which depends on the specific choice of
the waveguide geometry, is neglected and, for the sake of simplicity,
in the numerical examples we will use the free-space spontaneous decay
rate $\gamma_0=\omega_A^3 d_{eg}^2/(6\pi\epsilon_0\hbar c^3)$.

The waveguide modes form a one-dimensional continuum.  The above
equations can be transformed then into a much simpler form by means of
a Markoff approximation. On integrating Eq.~(\ref{eq:HL}a), the
waveguide field amplitudes arise in a form decomposed into a free
field part and into a part radiated by the atom \cite{cct},
\begin{equation}
  \label{eq:field_decomp}
  a_\omega(t) = a_\omega(t_0) + g_\omega e^{-i \omega z_A/c} 
  \int_{t_0}^t dt' \sigma_-(t') e^{i(\omega-\omega_A)t'}\; ,
\end{equation}
respectively. The back propagating modes are decomposed analogously.
The Markoff approximation is invoked to describe the back action of
the second term on the atom. It can be applied because the continuum
is broadband around the atomic frequency.  The free field term is
usually identified with a Langevin-type noise source.  In the
waveguide we can not make this step since we will consider initial
field states different from the vacuum. Altogether, the atom is
subject to the effect of the free field and to a relaxation into the
waveguide continuum. The polarization operator $\sigma_-$, for
example, follows
\begin{equation}
  \label{eq:sigma-}
  \frac{d}{dt} \sigma_- = -(\gamma_0+\gamma_1) \sigma_- +
     2\sigma_z \int d\omega g_\omega (a_\omega(t_0)
     e^{i\omega z_A/c}  + b_\omega(t_0) e^{-i\omega z_A/c}) 
     e^{-i(\omega-\omega_A) t} + \hat\xi_-\; ,
\end{equation}
where a new damping rate $\gamma_1$ appears, and the second term
contains the free field contributions. The vacuum frequency shift
induced by the waveguide modes, accompanying the relaxation, is
assumed to be already incorporated in a renormalized atomic frequency
$\omega_A$. The rate of spontaneous emission into the waveguide modes
is given by
\begin{equation}
  \label{eq:gamma1}
  \gamma_{1}=2\pi g_{\omega_A}^2 = \frac{1}{2} \frac{\sigma_A}{\cal A} \gamma_0 \; ,
\end{equation}
where the second expression directly exhibits the scaling with the
transverse extension of the waveguide. It is related to the atomic
radiative cross section $\sigma_A=3 \lambda^2/2\pi$.  A natural lower
bound on the transverse mode size is at about ${\cal A} \sim
(\lambda/2)^2$ (c.f. lowest order mode in box), implying a maximum
achievable coupling ratio $\gamma_1/\gamma_0 \sim 1$.  In the range
$\sigma_A \sim {\cal A}$, one has a strong waveguide-atom coupling,
which is manifested by the fact that the atom dissipates its energy
equally into the waveguide and the free-space ``lossy'' modes. This
situation could turn out to be a suitable basis to construct
single-atom detectors or efficient light emitting diodes.

For the following, it is convenient to introduce the ``free pulse'' operators
\begin{equation}
  g_{\omega_0} a_0(t) = \int_0^\infty d\omega g_\omega a_\omega(t_0)
  e^{-i(\omega-\omega_0) t}\; ,
\end{equation}
and the same for $b_0(t)$, where $\omega_0$ can be any frequency and
will later be identified with the central frequency of the wave
packet.  The evolution of the atomic operators is given by the
equations
\begin{subequations}
  \label{eq:Bloch_eff}
\begin{equation}
  \frac{d}{dt} \sigma_- = -\gamma \sigma_- +
    2\sigma_z g_{\omega_0} \left(a_0(t-z_A/c) e^{i\omega_0
    z_A/c} + b_0(t+z_A/c) e^{-i\omega_0 z_A/c} \right) 
    e^{-i(\omega_0-\omega_A) t} + \hat\xi_-
\end{equation}
\begin{multline}
  \frac{d}{dt} \sigma_z = -2\gamma (\sigma_z+1/2) -\\
    g_{\omega_0} \left(\sigma_+ 
    (a_0(t-z_A/c) e^{i\omega_0 z_A/c} + b_0(t+z_A/c) 
    e^{-i\omega_0 z_A/c}) 
    e^{-i(\omega_0-\omega_A) t } + h.c \right) + \hat\xi_z\; ,
\end{multline}
\end{subequations}
where $\gamma=\gamma_0+\gamma_1$ is the total decay rate of the atomic
dipole. This is the generalization of the Bloch equations for the
atomic dipole operators to describe the effect of a time-dependent
external excitation in the case of a quantum driving field. For
certain quantum states of the field, a closed set of equations can be
derived from the above operator equations to calculate the matrix
elements of the atomic operators. Note that only the free field
appears in the dynamics of the atomic dipole operators. This is a
direct consequence of the Markoff approximation, which implies that
any change of the electromagnetic field by the atom cannot act back
onto the atom. In other words, a photon emitted by the atom into the
waveguide leaves the interaction region immediately and cannot be
reabsorbed.  There is no feedback mechanism as provided by the mirror
in cavities.  Hence the strong coupling between atom and waveguide
yields a different dynamics with respect to the case of ordinary
cavity QED.

On inserting the solution of Eq.~(\ref{eq:Bloch_eff}a) into
(\ref{eq:field_decomp}), the field operator can be obtained
in the form
\begin{equation}
  \label{eq:field_terms_decomp}
  a_\omega (t) = a_\omega (t_0) +
  a_\omega^\sub{scat}(t) + a_\omega^\sub{back}(t) + 
   a_\omega^\sub{rad}(t) + a_\omega^\sub{noise}(t)\; ,
\end{equation}
where the five terms represent respectively the free field, the field
forward and backward scattered by the atom, the field radiated by an
initially excited atom, and finally the vacuum noise which is coupled
into the waveguide via the atomic dipole. They read
\begin{subequations}
  \label{eq:field_terms}
\begin{align}
  a_\omega^\sub{scat}(t) &= g_\omega g_{\omega_0} e^{-i
    (\omega-\omega_0) z_A/c} \int_0^t dt' e^{i(\omega-\omega_A+i\gamma) t'} 
    \int_0^{t'} dt'' e^{-i(\omega-\omega_A+i\gamma) t''} 2\sigma_z(t'')
    a_0(t''-z_A/c) \,, \\
  a_\omega^\sub{back}(t) &= g_\omega g_{\omega_0} e^{-i
    (\omega+\omega_0) z_A/c} \int_0^t dt' e^{i(\omega-\omega_A+i\gamma) t'} 
    \int_0^{t'} dt'' e^{-i(\omega-\omega_A+i\gamma) t''} 2\sigma_z(t'')
    b_0(t''+z_A/c) \,,\\ 
  a_\omega^\sub{rad}(t) &= g_\omega e^{-i \omega z_A/c} \int_0^t dt'
    e^{i(\omega-\omega_A+i\gamma) t'}\sigma_-(t_0) \,,\\
  a_\omega^\sub{noise}(t) &= g_\omega e^{-i \omega z_A/c} \int_0^t dt'
  e^{i(\omega-\omega_A+i\gamma) t'} \int_0^{t'} dt'' e^{\gamma t''}
    \hat\xi_-(t'')\; .
\end{align}
\end{subequations}
We have set the initial time $t_0=0$. Nevertheless, we still will use
$t_0$ sometimes in order to explicitly mark the initial value of the
operators evolving in the Heisenberg-picture.  In the double time
integrals the order can be exchanged to carry out one of them.
Substituting the solution in the electric field decomposition
(\ref{eq:E+}), the integral over the frequency $\omega$ can also be
performed. In this latter step, the lower bound of the frequency
integration is extended to $-\infty$. For $ct>z-z_A$, i.e.~a light
pulse has enough time to travel to the atom, one finds the
following result for the waveguide electric field in the half space
$z>z_A$
\begin{equation}
  \label{eq:E_result}
 \begin{split}
  E^{(+)}(\tau) = i & \sqrt{\frac{\hbar\omega_0}{4 \pi \epsilon c {\cal A}}} 
  e^{-i \omega_0 \tau} \Biggl[ a_0(\tau) +  \\
  &+\frac{1}{2} \frac{\sigma_A}{\cal A} \gamma_0 \int_0^{\tau+z_A/c}
  dt' e^{i(\omega_0-\omega_A+i\gamma) (\tau+z_A/c-t')} 2\sigma_z(t')
  a_0(t'-z_A/c) \\
  &+\frac{1}{2} \frac{\sigma_A}{\cal A} \gamma_0 e^{-2i\omega_0 z_A/c}
  \int_0^{\tau+z_A/c} dt' e^{i(\omega_0-\omega_A+i\gamma)
    (\tau+z_A/c-t')}
  2\sigma_z(t') b_0(t'+z_A/c) \Biggr]\\
  + i & \sqrt{\frac{\hbar\omega_A}{4 \pi \epsilon c {\cal A}}} 
  2\pi g_{\omega_A} e^{-(i\omega_A+\gamma)(\tau+z_A/c)} \Biggl[
  \sigma_-(t_0)+ \\
  &+ \int_0^{\tau+z_A/c} dt' e^{\gamma t'} \hat\xi_-(t') \Biggr]\; ,
\end{split}
\end{equation}
where $\tau=t-z/c$, and only the modes $a_\omega$ are considered as we
are interested in the field outgoing in the $+z$ direction ($z>z_A$).
In the subsequent lines one gets the free, the forward and the
backward scattered, and then the directly radiated and the noise
field. This is the central result that we use in the following to
calculate measurable quantities in various experimental scenarios. One
can draw a general conclusion already at this stage that the scattered
terms are proportional, as naively expected, to the ratio
$\sigma_A/{\cal A}$.

\section{Scattering of a single light pulse}

In this section we study the scattering process of a single light wave
packet off a ground state atom.  We calculate how the photons carried
by the wave packet are redistributed among the forward, backward and
transverse directions. The relevant quantity to be calculated is the
expectation value of the Poynting vector defined as
\begin{equation}
  \label{eq:Poynting_def}
  S(z,t) =  \frac{\cal A}{\hbar\omega_0} \frac{1}{\mu_0} 
  \langle \Psi_0 | B^{(-)} E^{(+)} + E^{(-)} B^{(+)}   | \Psi_0 \rangle\; ,
\end{equation}
given in units of photon current in the forward and backward direction
after the scattering event.  In the present scalar model the magnetic field
is just proportional to the electric field.  The initial state
$|\Psi_0\rangle$ is a simple product state of the waveguide field
state, the atomic ground state, and the environment:
\begin{equation}
  \label{eq:initial_state}
  |\Psi_0\rangle= |\psi_a\rangle|0_b\rangle|g\rangle|0_e\rangle\; ,
\end{equation}
The waveguide field describes a wave packet propagating in the
directions $+z$, i.e.~only the modes ${a_\omega}$ are excited with
state $ |\psi_a\rangle$, while counter propagating modes ${b_\omega}$
are in vacuum state $|0_b\rangle$. As for the initial state
$|\psi_a\rangle$ we will use Fourier-transform limited Gaussian
pulses in a coherent state or in a single-photon Fock state,
respectively. The mean Poynting vector of a free pulse is then
\begin{equation}
  \label{eq:free_Poynting}
  S_0(z,t)= N_a \frac{\Omega}{\sqrt{2 \pi}}
     e^{-\frac{1}{2} \Omega^2 (t-z/c)^2} \; ,
\end{equation}
where $N_a$ is the mean photon number carried by the wave packet and
$\Omega$ is the pulse bandwidth.

As no field is radiated from an atom in the
ground state, there is no contribution from the fourth term of
Eq.~(\ref{eq:E_result}) to the Poynting vector. Similarly the free space
environment as well as the backward propagating modes are in vacuum
state. Therefore, all the terms containing the noise operator ${\hat
  \xi}_-$ or $b_\omega(t_0)$ on the right-most side will vanish
acting on the initial state $| \Psi_0 \rangle$. Thus the mean Poynting
vector in the forward direction for $z>z_A$ is formed from the
superposition of the original light pulse and the one forward
scattered by the atom,
\begin{equation}
  \label{eq:Poynting}
 \begin{split}
 S(\tau) =& \frac{\Omega}{\sqrt{2 \pi}}
  e^{-\frac{1}{2} \Omega^2 \tau^2} \Biggl[
  \langle u(\tau)|u(\tau)\rangle + \\
  &+ \frac{\sigma_A}{\cal A} \gamma_0 
  \int_0^{\tau+z_A/c} dt' \Re{e\left\{{\cal F}(\tau,t')
  \langle u(\tau) | 2\sigma_z(t') | u(t'-z_A/c) \rangle\right\} } \\
  &+ \left(\frac{\sigma_A}{2\cal A} \gamma_0\right)^2 
  \int_0^{\tau+z_A/c} dt'  {\cal F}^*(\tau,t') \times\\
  & \qquad \int_0^{\tau+z_A/c} dt''  {\cal F}(\tau,t'') 
  \langle u(t'-z_A/c) |4\sigma_z(t') \sigma_z(t'')
  | u(t''-z_A/c) \rangle \Biggr] \; ,
\end{split}
\end{equation}
where $\tau=t-z/c$ and
\begin{equation}
 {\cal F}(\tau,t') = \exp{\left(\frac{\Omega^2}{4} (\tau+t'-z_A/c)
 (\tau + z_A/c -t') + i(\omega_0-\omega_A +i\gamma) (\tau + z_A/c -t')\right)}
  \; .
\end{equation}
The Poynting vector is expressed in a product form with the original
pulse shape function separated in a first term. To this end, we
defined an auxiliary state
\begin{equation}
  \label{eq:Psi}
  | u(t) \rangle = (2\pi\Omega^2)^{-\frac{1}{4}} e^{\frac{1}{4} \Omega^2
    t^2} {a}_0(t) | \Psi_0 \rangle \; .
\end{equation}
As we work in the Heisenberg picture, the time dependence of this
auxiliary state does not describe a dynamical process but follows from
its definition.

The reflected field in the domain $z<z_A$ results from the light
back-scattered by the atom (and additional quantum noise). As for the
atomic radiation the two directions in the waveguide are equivalent,
one gets an expression formally identical with the third term of
Eq.~(\ref{eq:Poynting}):
\begin{multline}
  \label{eq:Poynting_back}
 S^{\sub{(back)}}(\tau')=
  \frac{\Omega}{\sqrt{2 \pi}}e^{-\frac{1}{2} \Omega^2 \tau^2} 
 \left(\frac{\sigma_A}{2\cal A}\gamma_0\right)^2 \\  
 \int_0^{\tau'+z_A/c} dt' {\cal F}^*(\tau',t') 
  \quad \int_0^{\tau'+z_A/c} dt'' 
   {\cal F}(\tau',t'') \langle u(t'-z_A/c) |4\sigma_z(t') \sigma_z(t'')
  | u(t''-z_A/c) \rangle \; ,
\end{multline}
where $\tau'=t-(2z_A-z)/c$, expressing that the field propagates
now in the $-z$ direction.

\subsection{Coherent-state light pulse}

In the following we present how to evaluate the Poynting vector
(\ref{eq:Poynting}) in the case of wave packets initially in a coherent
state. A definition of multimode coherent state pulses can be found
in Ref.~\cite{Blow}. They have the property that they are
eigenstates of the annihilation operators. A coherent-state pulse of
bandwidth $\Omega$, centered initially at $z=0$, can be defined
through the eigenvalue equation
\begin{equation}
  \label{eq:coh_pulse}
  a_\omega (t_0) | \psi_a \rangle = \sqrt{N_a} \left(\frac{2}{\pi \Omega^2}
  \right)^{\frac{1}{4}} e^{-(\omega-\omega_0)^2/\Omega^2}  | \psi_a \rangle\;,
\end{equation}
where $N_a$ is the initial mean photon number in the wave packet. Its
photon statistics is Poissonian around the mean $N_a$ with variance
$\sqrt{N_a}$.

It follows that $|\Psi_0\rangle$ is an eigenstate of the free pulse
operator $ {a}_0(t)$ and we have:
\begin{equation}
 \label{eq:Psi_coh}
  |u(t)\rangle = \sqrt{N_a} |\Psi_0\rangle \; .
\end{equation} 
To evaluate the Poynting vector, Eq.~(\ref{eq:Poynting}), we need the
one-time and the two-time averages of the population inversion
operator. An equation of motion for $\langle\sigma_z(t)\rangle$ can be
directly obtained by taking the mean of the Eqs.
(\ref{eq:Bloch_eff}). Using again that $|\Psi_0\rangle$ is an
eigenstate of the free-pulse operators, the usual form of the optical
Bloch equations is found. We present it in the appendix. However,
when the evolution of $\langle\sigma_z(t)\sigma_z(t')\rangle$ as a
function of $t$ is calculated from (\ref{eq:Bloch_eff}), the operator
$\sigma_z(t')$ stands in between the free pulse operators and the
state $|\Psi_0\rangle$.  We prove in the same appendix that $\hat
a_0(t-z_A/c)$ and $\sigma(t')$ commute for $t>t'$ and therefore the
action of the pulse operator on its eigenstate can be easily carried
out.  This lemma allows the use of a form of the quantum regression theorem
to calculate the two-time average
$\langle\sigma_z(t)\sigma_z(t')\rangle$.

As it is shown in the appendix, an effective single-photon Rabi
frequency $g_\sub{eff}=g_{\omega_0} (2\pi \Omega^2)^{1/4}$ can be
identified in the Bloch equations. This coupling constant suggests
that the wave packet has an effective ``volume'' of $\sqrt{2\pi} c
{\cal A}/\Omega$. For larger bandwidths $\Omega$ we get a stronger
field per photon, which is a central quantity in cavity QED. However,
the interaction times between the atom and the transform limited
pulses get shorter in the same time, limiting the possibility of
coherent operations on the atomic states.

The photon number $N_a$ multiplies all the terms in the Poynting
vector (\ref{eq:Poynting}).  Beyond this simple linear dependence, the
photon number has an influence on the evolution of the atomic
population. Through this term, the atomic saturation introduces a
nonlinear behavior into the Poynting vector, which can be a noticeable
effect in the strong coupling regime even for relatively small photon
numbers $N_a$.

\subsection{Fock-state light pulse}

Now lets turn to a second example and study a wave packet of precisely
one photon.  The single-photon Fock-state is defined by
\begin{equation}
  \label{eq:1Fock_def}
  | 1_a\rangle = \int_0^\infty d\omega \,
    \left(\frac{2}{\pi\Omega^2}\right)^{\frac{1}{4}}
    e^{-(\omega-\omega_0)^2/\Omega^2} {a}_\omega^\dagger (t_0)
    |0_a\rangle \; .
\end{equation}
The definition in (\ref{eq:Psi}) leads to
\begin{equation}
 \label{eq:Psi_fok}
  |u(t)\rangle = |0_a\rangle |0_b\rangle |g\rangle |0_e\rangle\; ,
\end{equation}
which, similarly to the coherent-state case Eq.~(\ref{eq:Psi_coh}), is
time-independent.  We now need the diagonal matrix elements $\langle
0, g | \sigma_z(t) |0, g \rangle$ and $\langle 0,
g|\sigma_z(t)\sigma_z(t')|0, g\rangle$ where $0$ stands for both modes
$a$ and $b$ and for the environment.  Equation (\ref{eq:Bloch_eff})
implies that $\langle 0, g | \sigma_z(t) |0, g \rangle=-1/2$ for
arbitrary time $t$. In fact, this value is the initial condition
itself that makes the right hand side of the generalized Bloch
equations vanish.  Using again the commutation of the operators $\hat
a_0(t-z_A/c)$ and $\sigma(t')$, one can show that $\langle 0,
g|\sigma_z(t)\sigma_z(t')|0, g\rangle=1/4$ independently of $t$ and
$t'$.  We get such simple solutions because we consider, formally, the
Bloch equations when the atom is driven by a ``vacuum pulse''.

Having the exact solutions for the one-time and two-time averages, the
integrals in Eq.~(\ref{eq:Poynting}) can be analytically evaluated.
The integration invokes the complex error function and the result
itself is not very instructive. We omit this rather long
formula here and only show typical results graphically in the figures
of the next section.

\subsection{Single-atom transmittance and reflectance}

We continue the discussion of light pulse scattering on a single atom
by means of numerical examples. For this we evaluate the Poynting
vector expressions (\ref{eq:Poynting}) and (\ref{eq:Poynting_back}) to
determine how much of a light pulse is transmitted and reflected back
by the atom. We introduce the transmittance, i.e., the transmitted
mean energy (photon number) divided by the mean energy of the
impinging wave packet. For an atom being on resonance with the pulse
carrier frequency $\omega_A=\omega_0$ we expect the strongest
scattering. In the forthcoming examples the waveguide cross section is
set to $\sigma_A={\cal A}$.

\begin{figure}[h]
\includegraphics*[width=9cm]{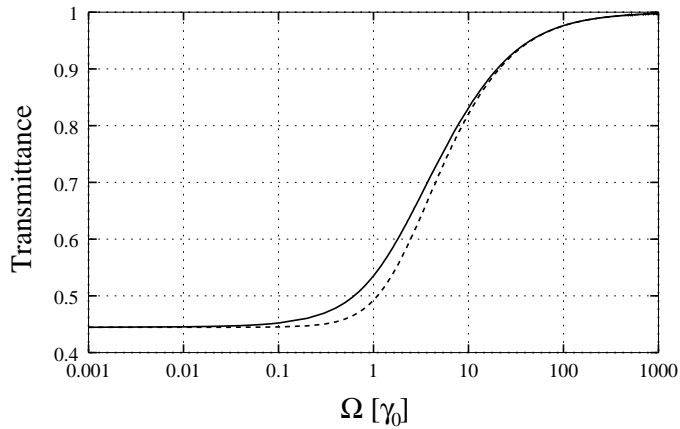}
\caption{Resonant single-atom transmittance as a function of the pulse
  bandwidth for pulses initially in coherent state with mean photon
  number $N_a=1$ (solid line) and in single-photon Fock state (dashed
  line).  The parameters are $\omega_A=\omega_0$ (resonant pulse) and
  $\sigma_A={\cal A}$. }
\label{fig:transmittance}
\end{figure}

In Figure \ref{fig:transmittance} the single-atom transmittance is presented
as a function of the pulse bandwidth on a logarithmic scale.
The overall shape of the transmittance curve is very similar for a
pulse initially in a coherent state with $N_a=1$ (solid line) or if it
is in a single-photon Fock state (dashed line). The difference is
manifest for bandwidths close to the transition linewidth,
$\Omega\approx\gamma_0$. Apparently the effective
saturation is reduced if exactly one photon is present instead of a
distribution with mean one.
 
As the incoming mean photon number is now 1 in this example, the
transmittance gives directly the mean transmitted photon number. The
atom presents an ``obstacle'' provided the pulse length is on the
order of the natural lifetime or longer, i.e., for $\Omega<\gamma_0$. The
single-atom effect on the transmittance is approximately constant up
to pulse lengths of few times the atomic lifetime. In this range the
transmitted power is less than the half of the incoming one. The lower
bound in the monochromatic limit is given by
\begin{equation}
  {\cal T}=
  1-2\frac{\gamma_1}{\gamma}+\left(\frac{\gamma_1}{\gamma}\right)^2 \; ,
\end{equation}
which is $4/9\approx 0.44$ in the example. There is an intermediate
regime $\gamma_0 < \Omega < 10 \gamma_0$ where the transmittance
changes rapidly and it is noticeably smaller for a Fock-state pulse than
for a coherent state. Finally, short light pulses, $\Omega \gg \gamma_0$, pass
through the atom almost without being appreciably affected, realizing
a sort of short pulse filter. A simple explanation is
that a large spectral part of a broad-band pulse is far away
from the atomic resonance, which reduces the effective coupling strength.
Figure \ref{fig:transmittance} therefore suggests that in situations where
temporal resolution is required, for
example in a detection scheme, the best choice for the bandwidth is
$\Omega\approx\gamma_0$ where relatively short light pulses still experience
significant attenuation. Longer pulses provide less temporal resolution
without much improvement of the signal, whereas shorter pulses experience a
significantly smaller change.

Let us note briefly that the part of the impinging pulse which is not
transmitted is either reflected or, most probably, scattered in the
transverse lossy modes. The reflectance has an upper bound which is
reached again in the monochromatic limit. It is
$(\gamma_1/\gamma)^2=1/9 \approx 0.11$ in the numerical example
considered for Fig.~1, that is at least with probability 4/9 the
photon is lost from the waveguide.

\begin{figure}[htbp]
  \begin{center}
    \includegraphics*[width=9cm]{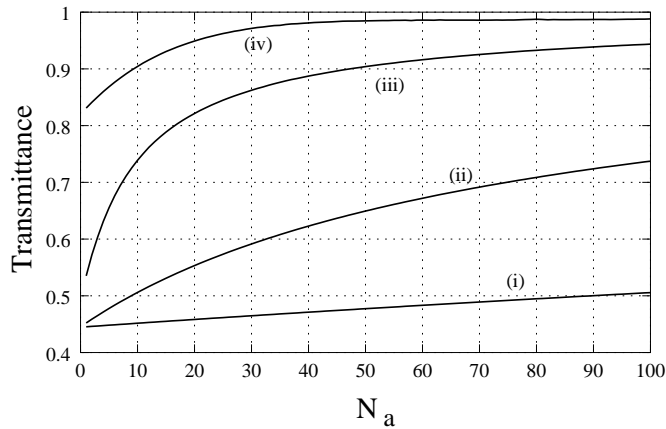}
    \caption{The transmittance as a function of the mean photon number
      $N_a$ of the initial coherent-state pulse. Curves (i), (ii),
      (iii), and (iv) are associated with the pulse bandwidths $\Omega=
      \gamma_0/100$, $\gamma_0/10$, $\gamma_0$, and $10 \gamma_0$,
      respectively.}
    \label{fig:nonlinear}
  \end{center}
\end{figure}

Due to saturation, there is a nonlinear effect in the transmittance.
The more photons the light pulse carries, the more the atom gets
excited. Population in the upper state of the atom reduces effectively
the dipole strength, as follows from the term $\langle \sigma_z
\rangle$ in the Eq.~(\ref{eq:Poynting}), which amounts therefore to a
decrease in the amount of scattering.  This effect is exhibited in
Figure \ref{fig:nonlinear} where the transmittance is plotted against
the mean photon number $N_a$ of the initial coherent-state pulse for
various pulse bandwidths. Parameters are the same as in
Fig.~\ref{fig:transmittance}.
The starting points of the curves at $N_a=1$ correspond to four
different points on the curve of the previous figure
\ref{fig:transmittance}.  For longer pulses, see the curve (i) or (ii)
in the figure, the saturation effect is small. This is because
the effective coupling strength $g_\sub{eff}$ decreases with the pulse
length. Or, alternatively, one can view the same effect as a
consequence of the photons arriving more distributed in time. For
short pulses, on the other hand, the coupling constant becomes large
enough to reach saturation even with weak incident light pulses. For
example, curve (iii) corresponding to $\Omega=\gamma_0$ manifests a
drastic nonlinear behavior even for photon numbers around 1.

\begin{figure}[htbp]
  \begin{center}
    \includegraphics*[width=14cm]{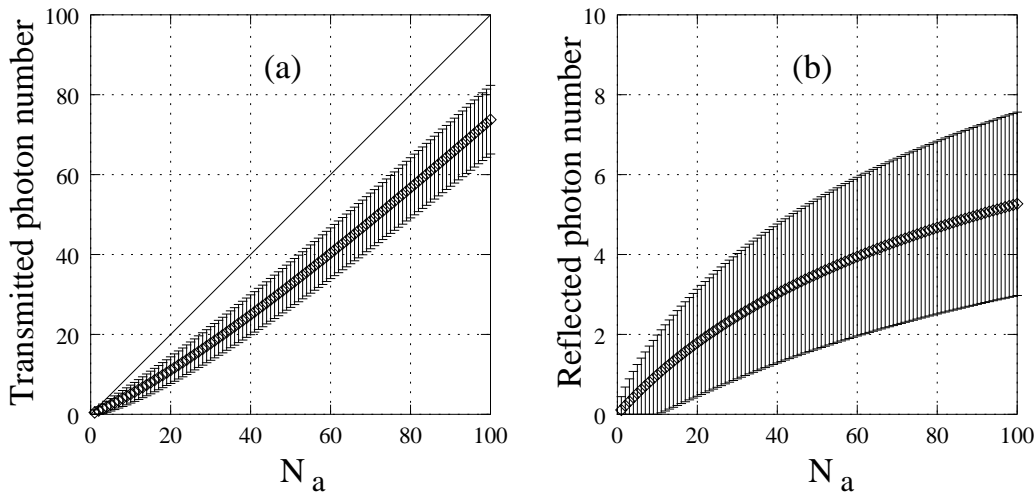}
    \caption{Transmitted (a) and reflected (b) number of photons as a
      function of the initial mean photon number $N_a$. Error bars
      indicate the square root of the mean signal corresponding
      to the noise in a coherent state. The bandwidth is set
      $\Omega=0.1 \gamma_0$.}
    \label{fig:detection}
  \end{center}
\end{figure}

Figure \ref{fig:detection} shows transmitted and reflected photon numbers for
$\Omega=\gamma_0/10$ (same as in curve (ii) of figure
\ref{fig:nonlinear}) depending on the mean photon number of the pulse. The
square root of the mean photon number is shown with error bars,
representing the inherent quantum noise associated with coherent
states.
In order to detect the single-atom scatterer with high probability,
the change in the photon number (reflected or transmitted) must be
larger than the noise.  The figure reveals the minimum necessary
photon number of the probe pulse to achieve the required operation
regime. Photon numbers at about 20 in the probe pulse are sufficient
to get a high quantum efficiency for the detector.

\subsection{Phase shift and pulse deformation}

When the carrier frequency is detuned from atomic resonance, the
pulse can undergo a phase shift (dispersive scattering). This phase
shift can be the basis of a single-atom birefringence in waveguides.
This happens if the atomic dipole is parallel to one of the
polarization modes and orthogonal to the other one. Only the first
polarization mode experiences a phase shift whose magnitude will be
estimated, based on Eq.~(\ref{eq:E_result}), in this section.

Strictly speaking the study of the phase properties necessitates the
definition of a phase operator and the investigation of its time evolution.
Here we will adopt a
simplified treatment which allows perhaps a more instructive insight
in the phase properties. Initially a coherent state pulse, defined in
(\ref{eq:coh_pulse}), is taken with the phase 0, i.e., the amplitude
of all the components is real. We assume that the state stays almost a
coherent state all along the evolution. This is fulfilled when the
pulse induces little atomic population excitation.  Then, formally,
the inversion operator $\sigma_z$ can be replaced by $-1/2$ times the
unity operator, and the state $| \Psi_0 \rangle$ is indeed an eigenstate of the
electric field operator, Eq.~(\ref{eq:E_result}), at any time.
The phase can be described by the argument of the corresponding complex
eigenvalue, which we can write as
\begin{equation}
  \label{eq:eigen_psi0}
E^{(+)}(\tau) |\Psi_0 \rangle = i \sqrt{\frac{\hbar\omega}{4 \pi
  \epsilon c {\cal A}}}  
  e^{-i \omega_0 \tau} \alpha_0(\tau)\left( 1 + h(\tau)
  \right) | \Psi_0 \rangle \; ,
\end{equation}
where $\alpha_0(\tau)=\sqrt{N_a} (2\pi\Omega^2)^{1/4}
\exp{(-\Omega^2\tau^2/4)}$ describes the initial Gaussian pulse shape.
The radiated and the backscattered terms, the third and the fourth
terms in Eq.~(\ref{eq:E_result}), vanish because of the choice of the
initial condition. The noise term is negligible as it is proportional
to the atomic excitation, which is supposed to be small. Thus, the
term $h(\tau)$ stems exclusively from forward scattering.
It can be generated in a simple form by changing the integration
variable in Eq.~(\ref{eq:E_result}) as $t'\rightarrow \tau+z_A/c-t'$.
One gets
\begin{equation}
  \label{eq:phase}
 h(\tau) =  -\frac{1}{2} \frac{\sigma_A}{\cal A}
  \gamma_0 \int_{0}^{\tau+z_A/c} dt' e^{i(\omega_0-\omega_A)t'}
  e^{-(\gamma + 2 \Omega^2\tau) t'} e^{-\frac{1}{4}
  \Omega^2 {t'}^2}\; .
\end{equation}
On inspecting Eqs.~(\ref{eq:eigen_psi0}) and (\ref{eq:phase}), two
effects of the forward scattering on the coherent-state amplitude can
be noticed. First of all, the coherent-state amplitude is shifted 
with a relative amount of $h(\tau)$. Second, this shift depends on
$\tau=t-z/c$ which yields a deformation with respect to the initial
Gaussian pulse shape.  The upper integration bound in
Eq.~(\ref{eq:phase}) can be extended to infinity if the
distance $z_A$ between the initial pulse and the atom is very large
compared to the pulse extension $c/\Omega$.  Hence, the only
dependence on $\tau$ derives from the term $\gamma + 2 \Omega^2\tau$
in the exponent. As $\tau$ varies within $\pm 1/\Omega$ in the pulse,
the ratio of $\gamma$ and $\Omega$ decides how much the pulse shape is
distorted.

For very long pulses (quasi-monochromatic excitation limit) the
exponential decay with $\gamma$ dominates the integrand. The other
exponentials can be neglected and the integral renders the well-known
Lorentzian form of the single-atom susceptibility $\chi= (\sigma_A/2
{\cal A})/( (\omega_0-\omega_A) +i\gamma)$. Evidently, in this limit
the result does not depend on $\tau$. For general relation
between $\Omega$ and $\gamma$, the integral has to be evaluated
numerically. Let us first study the change of the coherent-state
amplitude as a function of the detuning, and then its dependence on
$\tau$.

\begin{figure}[htbp]
  \begin{center}
    \includegraphics*[width=9cm]{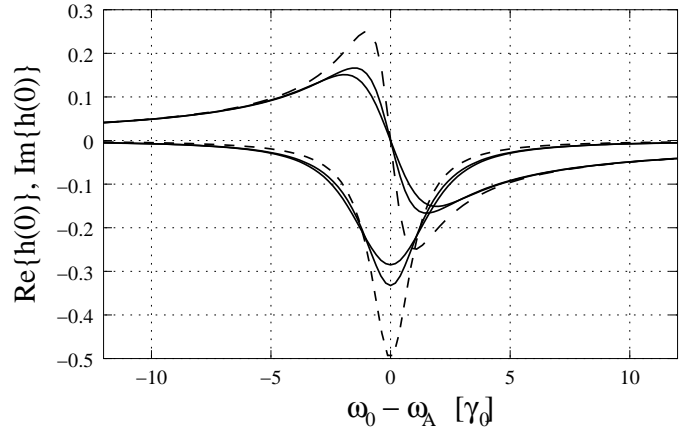}
    \caption{Change of the coherent-state amplitude as a function of the pulse
      carrier frequency detuning from the atomic resonance. Both the
      ``absorption'' $\Re{e\{h(0)\}}$ and the ``dispersion''
      $\Im{m\{h(0)\}}$ curves are plotted with solid lines for
      $\Omega=\gamma_0$ and $\Omega=0.1 \gamma_0$. This latter
      corresponds to the larger $h(0)$ values.  Dashed curves
      show the Lorentzian shape for reference. }
    \label{fig:susceptibility}
  \end{center}
\end{figure}

For the first case, we fix $\tau=0$ and plot the numerical solution of
$h(\tau=0)$ in Figure \ref{fig:susceptibility}.  We show the real and
the imaginary parts of $h(0)$ as a function of the detuning
$\omega_0-\omega_A$ for two bandwidth values, $\Omega=\gamma_0$ and
$\Omega=0.1 \gamma_0$. By analogy with the susceptibility, the real
and imaginary parts can be regarded as an absorption and a
dispersion-like curve, respectively. For reference, the Lorentzian
function associated with the quasi-monochromatic excitation limit
(linewidth $\gamma$, oscillator strength $\sigma_A/2 {\cal A}$) is
also plotted with dashed lines.
The three curves merge asymptotically for very large detunings. The
reason is that the frequency components of the pulse are more or less
uniformly detuned from the atomic frequency, much like in the
monochromatic limit.  Hence, the phase shift can be well approximated
from the analytic Lorentzian solution. Otherwise, for moderate
detunings, the bandwidth dependence of the change in the
coherent-state amplitude is quite apparent in the figure. The narrower
the bandwidth, the larger phase shift is obtained.

\begin{figure}[htbp]
  \begin{center}
    \includegraphics*[width=9cm]{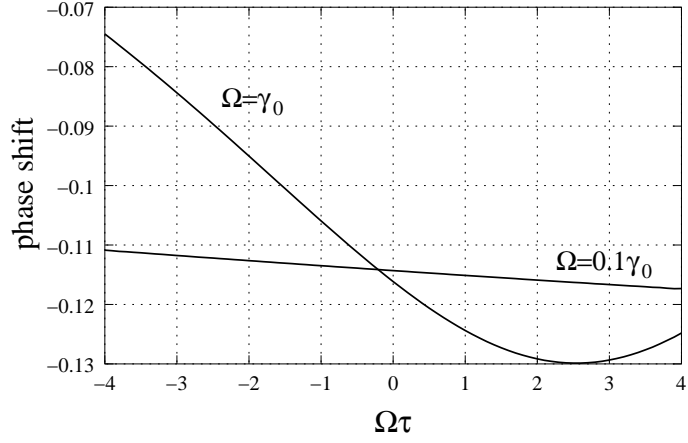}
    \caption{The phase shift as a function of the pulse position
      $\Omega \tau$. For narrow enough bandwidth the phase shift gets
      independent of $\tau$. Parameters are 
      $\omega_0-\omega_A=4 \gamma_0$, $\sigma_A/{\cal A}=1$.}
    \label{fig:phaseshift}
  \end{center}
\end{figure}

When the change in the coherent-state amplitude $h(\tau)$ is small
with respect to 1 and approximately proportional to $i$, that is, for
$\omega_0-\omega_A\gg \gamma_0$, it
directly gives the rotation angle in phase space, i.~e., the phase
shift.  The phase shift as a function of $\tau$ is presented in Figure
\ref{fig:phaseshift}.
The detuning is chosen $\omega_0-\omega_A = 4 \gamma_0$ where the
absorption is well reduced and the pulse power is almost completely
transmitted through the atom. One can see in the figure that rotation
angles in the range of 0.11 rad$\approx 6^o$ are induced by a single
atom for $\sigma_A={\cal A}$.  Note that for $\Omega=0.1\gamma_0$ the
phase shift is almost uniform, while it changes considerably for
$\Omega=\gamma_0$. In both cases the deformation is asymmetric,
showing that the front and the back of the pulse interact with an atom
in a different state.

\section{Interference of light pulses}

One possible application of such a setup is nonlinear photonics, i.e.
doing optics with few photons beyond beam splitters and mirrors
\cite{resch}. As a prototype of this dynamics let us study now the
``collision'' of two pulses propagating in opposite directions.
Instead of considering the modes $b_\omega(t_0)$ initially in vacuum
state, we will properly define an initial state that describes a wave
packet centered at $z=2z_A$. Then, this backward propagating wave
packet and the forward propagating one encounter in the position of
the atom at $z=z_A$ where the atomic dipole can mediate an interaction
between them.  The calculations are based on the general result
presented in Eq.~(\ref{eq:E_result}), which holds for any initial
condition.

We calculate the Poynting vector in the range $z > z_A$ after the
pulse collision.  Here the field is composed of the initial pulse in
the modes $a_\omega$ and its forward scattered part, superimposed with
the back scattered part of the pulse in the modes $b_\omega$. One gets
an expression quite similar to the result in Eq.~(\ref{eq:Poynting}),
\begin{equation}
  \label{eq:Poynting_coll}
 \begin{split}
  S(\tau) =& \frac{\Omega}{\sqrt{2 \pi}}
  e^{-\frac{1}{2} \Omega^2 \tau^2} \Biggl[
  \langle u(\tau)|u(\tau)\rangle + \\
  &+ \frac{\sigma_A}{2 \cal A} \gamma_0 
  \int_0^{\tau+z_A/c} dt'  2\Re{e\left\{{\cal F}(\tau,t')
  \langle u(\tau) | 2\sigma_z(t') | v(t'-z_A/c) \rangle\right\}} \\
  &+ \left(\frac{\sigma_A}{2\cal
  A} \gamma_0\right)^2 \int_0^{\tau+z_A/c} dt'  {\cal F}^*(\tau,t') \\
  &\qquad \int_0^{\tau+z_A/c} dt'' {\cal F}(\tau,t'')
   \langle v(t'-z_A/c) |4\sigma_z(t') \sigma_z(t'')
  | v(t''-z_A/c) \rangle \Biggr] \; ,
\end{split}
\end{equation}
just we needed to introduce a second state $|v (t)\rangle$, which
is defined by
\begin{equation}
  \label{eq:Phi}
  | v(t) \rangle =  (2\pi\Omega^2)^{-\frac{1}{4}} e^{\frac{1}{4} \Omega^2
    t^2} \left({a}_0(t) + e^{2i\omega_0 z_A/c}
    b_0(t+2z_A/c) \right) | \Psi_0 \rangle \; .
\end{equation}

\begin{figure}[htbp]
  \begin{center}
    \includegraphics*[width=9cm]{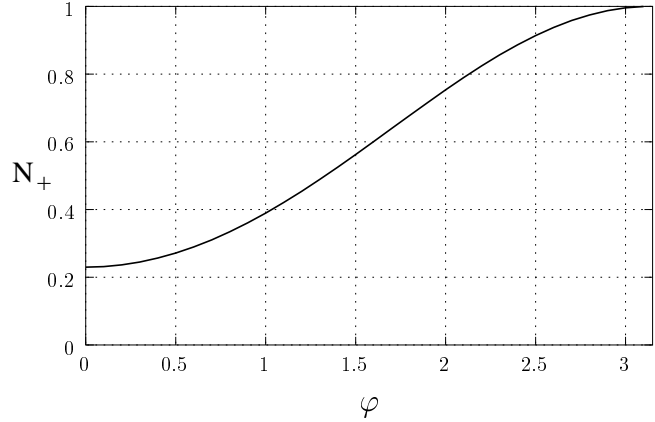}
    \caption{Mean photon number $N_+$ 
      outgoing in the $+z$ direction as a function of the relative
      phase $\varphi$ of two identical counter propagating
      coherent-state pulses. The mean photon numbers are $N_a=N_b=1$,
      the bandwidth is $\Omega=0.3 \gamma_0$, and $\sigma_A/{\cal
        A}=1$.}
    \label{fig:fringes}
  \end{center}
\end{figure}

Let us consider first the case when the two counter-propagating
pulses are in a coherent state. The initial state is composed of the
state $|\psi_a\rangle$ as defined in Eq.~(\ref{eq:coh_pulse}), while 
the back-propagating modes are in the state $| \psi_b\rangle$ where 
\begin{align}
  \label{eq:coh_pulse2}
  b_\omega (t_0) | \psi_b \rangle = \sqrt{N_b} e^{i\varphi}
  \left(\frac{2}{\pi \Omega^2} \right)^{\frac{1}{4}}
  e^{-(\omega-\omega_0)^2/\Omega^2} e^{2i\omega z_A/c} | \psi_b \rangle \; .
\end{align}
The last exponential term ensures that the position of the pulse is
$z=2z_A$ at $t=t_0$. The state $ | v(t) \rangle$ is proportional
again to the initial state,
\begin{equation}
 \label{eq:Phi_coh}
  |v(t)\rangle = \left(\sqrt{N_a}+e^{i\varphi}\sqrt{N_b}\right) 
  |\Psi_0\rangle \; .
\end{equation} 
One can immediately recognize that the appearance of the state $|v(t)
\rangle$ in the second and third terms of Eq.~(\ref{eq:Poynting_coll})
gives rise to intriguing interferometric phenomena. Compared to the
simple scattering scenario given by Eq.~(\ref{eq:Poynting}), the
second term, which is responsible for the absorption, can be enhanced.
To this end, the counter-propagating pulses must have the same phase,
i.e.~$\varphi=0$. Then the forward scattered pulse and the
backscattered pulse interfere constructively yielding a reduced
transmitted signal.  On the other hand, the two pulses can interfere
destructively when they black out each other in the position of the
scatterer, i.e.\ for $N_a=N_b$ and $\varphi=\pi$.  The atom being in
dark, this situation looks as if the atom were absent and the pulses
propagate freely. This simple example demonstrates that there is
indeed a non-trivial interaction between the pulses mediated by a
single atom.  Since the transverse scattering is missing, the outcome
is definitely different from the coherent sum of the outcomes obtained
with two single pulses. In between the extreme cases the outgoing mean
photon number in the $+z$ direction, denoted by $N_+$ varies as a
function of the phase $\varphi$. This is shown in
Fig.~\ref{fig:fringes} where a difference of almost 80\% is obtained
for a bandwidth $\Omega=0.3 \gamma_0$ and a beam cross section
$\sigma_A/{\cal A}=1$.

\begin{figure}[htbp]
  \begin{center}
    \includegraphics*[width=9cm]{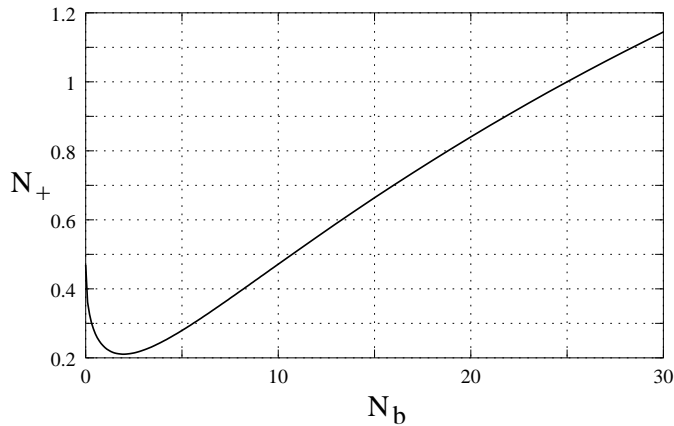}
    \caption{Mean photon number $N_+$ 
      outgoing in the $+z$ direction as a function of the mean photon
      number $N_b$ of the pulse incoming from the backward $-z$
      direction. Parameters are as in figure \ref{fig:fringes}.}
    \label{fig:tuning}
  \end{center}
\end{figure}

Assume that we have a mean photon number $N_a$ fixed, e.g.~$N_a=1$.
Then the photon number $N_+$ outgoing in the forward $+z$ direction
can be tuned by varying the photon number $N_b$. This effect is
represented in the Figure \ref{fig:tuning}.
The well-pronounced minimum results from the concurrence of the second term of
Eq.~(\ref{eq:Poynting_coll}), which scales as $\sqrt{N_b}$ and reduces the
outgoing photon number, and the third term, which is linear in $N_b$ and
increases the photon flux due to back scattered light. The manifest deviation
from the linear behavior is due to atomic saturation.

Let us finally consider the collision of two Fock-state pulses on the
atom. The initial state is $|\Psi_0\rangle = | 1_a\rangle |1_b\rangle 
|g\rangle |0_e\rangle$, where $|1_b\rangle$ is defined by
\begin{equation}
  \label{eq:11Fock_def}
  | 1_b\rangle = 
   \int_0^\infty d\omega \,
    \left(\frac{2}{\pi\Omega^2}\right)^{\frac{1}{4}}
    e^{-(\omega-\omega_0)^2/\Omega^2} e^{2i\omega z_A/c} 
   {b}_\omega^\dagger (t_0)
    |0_b\rangle \; .
\end{equation}
Initially there is no quantum correlation between the two Fock-state
wave packets. As the phase in Fock states is completely undefined, one
expects that the interference is missing in this case. This is true
and can formally traced back to the fact that for the auxiliary state
of Eq.~(\ref{eq:Poynting_coll}),
\begin{equation}
 \label{eq:Phi_fok}
  |v(t)\rangle = |0_a,1_b\rangle + |1_a,0_b\rangle\; ,
\end{equation} 
the contribution of ${a}_0(t'-z_A/c)$ (``forward scattered photon'')
and the contribution of ${b}_0(t'+z_A/c)$ (``back reflected photon'')
do not add up algebraically. Note that for coherent-state pulses this
was different, the second ``absorption'' term of
Eq.~(\ref{eq:Poynting_coll}) got a factor
$\sqrt{N_a}+e^{i\varphi}\sqrt{N_b}$ and the Bloch equations had to be
solved only for the mean of atomic operators. Here, for Fock states,
one has to solve a specific realization of the generalized Bloch
equations (see in the Appendix). However, these equations are
inhomogeneous and the solution is far from being the sum of two terms,
one originating in $|0_a,1_b\rangle$ and the other in
$|1_a,0_b\rangle$. The numerically obtained result for the mean photon
number outgoing in the $+z$ direction is very similar, however, to the one
obtained with one single impinging pulse. No interferometric
enhancement or reduction of the outgoing photon number can be observed
for pulses initially in Fock states.

\section{Conclusion}

For a field transversally confined in a tiny waveguide a single atom
is able to have a significant effect on a light wave packet traveling
across the atom.  As limiting cases of the time-dependent 1D
scattering, one finds a transmittance reduction below 50\%, strong
nonlinearity in the transmittance even for very low energy pulses, and
interferometric coupling of two wave packets with a visibility of up
to 80\%.  All these effects clearly demonstrate that miniaturized
waveguides or fibers coupled to an atomic dipole can be used as
efficient single-atom detectors, or photon--photon couplers.  As one
might be able to control a single atom state on the quantum level,
this could pave the way to genuine quantum devices for optics, as
e.g.~a quantum switch for light \cite{Gheri}. Similarly one could
envisage to construct single photon Bell state analyzers, as they
would be needed for improved quantum cryptography or quantum
teleportation setups.

In our model we aimed to studying the fundamental nature of the
interaction of a single atom with a waveguide field. A regime that can
be referred to as ``strong coupling regime'' occurs when the
transverse extension of the waveguide modes is close to the
single-atom radiative cross section. The coherent interaction between
the field and the atom can dominate damping. In contrast to cavity
QED, however, the waveguide-atom coupling yields a dissipation-like
evolution of the atom since the waveguide modes form a broadband
continuum, similarly to the reservoir composed of the free-space
radiation modes. On one hand, a single atom within the waveguide field
becomes a significant scatterer, as featured by the effects presented
in the paper. On the other hand, it is questionable if a weak quantum
field can perform coherent population transfer in an atom. This is
because a large effective single-atom Rabi frequency ($g_\sub{eff}$)
is accompanied by the presence of a large damping rate ($\gamma_1$) in
the Bloch equations.  In addition to this, the Rabi frequency depends
on the interaction time defined by the pulse length. For example, for
adiabatic passage techniques long interaction time is required, which
inevitably leads to the reduction of the coupling constant. Our model
equations provide a suitable ground to further study the dynamics of
atoms coupled to one-dimensional continua of modes.

\acknowledgments

We thank J\"org Schmiedmayer and Ron Folman for motivating this
research by their interest in constructing single-atom detectors. We
also thank Peter Zoller for stimulating discussions.  This work was
supported by the Austrian Science Foundation FWF (Project P13435).
P.~D.\ acknowledges the financial support by the National Scientific
Fund of Hungary under contracts No. T023777 and F032341.

\appendix

\section{Derivation of various matrix elements of the atomic operators}

We prove in this appendix that the free pulse operator $a_0(t-z_A/c)$
commutes with the population inversion operator $\sigma_z(t')$ for
times $t>t'$. This lemma allows the derivation of a closed set of
linear differential equations for the required matrix elements of the
population inversion operator and for that of the two-time product
$\sigma_z(t)\sigma_z(t')$. We briefly present these equations which
are formally very similar to the Bloch equations and the ones obtained
by the quantum regression theorem in the case of driving an atomic
dipole transition with a classical field.

By definition,
\begin{multline}
  \label{eq:commutator}
  C \equiv \left[a_0(t_1-z_A/c),\sigma_z(t_2)\right] = 
  \int d\omega e^{-i(\omega-\omega_0)(t_1-z_A/c)} \left[\hat
  a_\omega(t_0), \sigma_z(t_2)\right]\\
  = \int d\omega e^{-i(\omega-\omega_0)(t_1-z_A/c)} \left\{ \left[\hat
  a_\omega(t_2), \sigma_z(t_2)\right] - g_\omega e^{-i\omega z_A/c} 
  \int_0^{t_2} dt' e^{i(\omega-\omega_A) t'}\left[\sigma_-(t'),
  \sigma_z(t_2)\right] \right\}\; ,
\end{multline}
where in the second step we used Eq.~(\ref{eq:field_decomp}). Atomic
and field operators, taken at the same time, commute, hence the first
term vanishes. In the second term one can change the order of the
integrals
\begin{equation}
  C= -e^{-i \omega_0 z_A/c} \int_0^{t_2} dt' e^{i(\omega_0-\omega_A)
  t'} \left[\sigma_-(t'), \sigma_z(t_2)\right] \int d\omega g_\omega
  e^{-i(\omega-\omega_0)(t_1-t')}  \; .
\end{equation}
The coupling constant $g_\omega$ being a slowly varying function of
$\omega$ around the given optical frequency $\omega_0$ can be taken
out of the integral. The remaining inner integral amounts to a
Dirac-delta function and 
\begin{multline}
  C = -e^{-i \omega_0 z_A/c} \int_0^{t_2} dt' e^{i(\omega_0-\omega_A)
    t'} \left[\sigma_-(t'), \sigma_z(t_2)\right]\, 2\pi g_{\omega_0} 
  \delta(t'-t_1)= \\
  = - 2\pi g_{\omega_0} e^{-i \omega_0 z_A/c} e^{i(\omega_0-\omega_A)
    t_1} \left[\sigma_-(t_1), \sigma_z(t_2)\right] \Theta(t_2-t_1)\; ,
\end{multline}
where $\Theta$ is the Heavyside step function, which proves our conjecture.

The time dependent effect of the light pulse on the atomic dipole
operators can be determined from the Eq.~(\ref{eq:Bloch_eff}).
Depending on the initial state of the pulses, various matrix elements
have to be calculated and inserted in the expressions for the Poynting
vector Eqs.~(\ref{eq:Poynting}), (\ref{eq:Poynting_back}) and
(\ref{eq:Poynting_coll}). In all the present cases, the required
variables obey a set of linear differential equations that can be
written in the form
\begin{equation}
 \label{eq:Bloch}
\dot{ {\bf s}} = {\bf B} {\bf s} +{\bf b} \; .
\end{equation}

When the initial pulses are in coherent states, the evolution of the
quantum mean of the population inversion operator is needed. It can be
obtained by integrating the above differential equation with
\begin{multline}
{\bf s}= \left( \begin{array}{c}
 \langle \hat\sigma_z \rangle\\
 \Re{e\left\{\langle \hat\sigma'_+ \rangle \right\} }\\
 \Im{m\left\{\langle \hat\sigma'_+ \rangle \right\} } 
 \end{array} \right)\; ,\quad 
{\bf b} = \left( \begin{array}{c} 
-\gamma \\ 
0 \\ 
0 \end{array} \right)\; , \\
{\bf B}= \left( \begin{array}{ccc}
-2 \gamma & - 2 (\sqrt{N_a}+\sqrt{N_b}\cos{\varphi}) g(t) &
2 \sqrt{N_b}\sin{\varphi} g(t)\\
2 (\sqrt{N_a}+\sqrt{N_b}\cos{\varphi}) g(t) & - \gamma & \omega_0-\omega_A  \\ 
-2 \sqrt{N_b}\sin{\varphi} g(t) & \omega_A-\omega_0 & -\gamma \end{array} \right)\; ,
\end{multline}
where
\begin{equation}
\hat\sigma'_+ (t)= \hat\sigma_+(t) e^{i\omega_0 z_A/c}
    e^{-i(\omega_0-\omega_A) t} 
\end{equation}
Formally, this equation is equivalent with the Bloch equations obtained
for an atom driven by a classical time-dependent excitation. The
corresponding single-photon Rabi frequency reads 
\begin{equation}
  g(t) = g_{\sub{eff}} e^{-\frac{1}{4} \Omega^2 (t-z_A/c)^2}\; ,
\end{equation}
with $g_{\sub{eff}}=g_{\omega_0} (2 \pi \Omega^2)^{1/4}$.

The two-time averages in Eq.~(\ref{eq:Poynting}) can be obtained by using
the quantum regression theorem. For a fixed time $t'$
prior to time $t$, the variables
\begin{equation}
 \label{eq:QST}
{\bf s} (t) = \left( \begin{array}{c}
 \langle \hat\sigma_z(t) \hat\sigma_z(t')\rangle\\
 \left\langle \Re{e\left\{ \hat\sigma_+'(t) \right\} } \hat\sigma_z(t')\right\rangle\\
 \left\langle \Im{m\left\{ \hat\sigma_+'(t) \right\} } \hat\sigma_z(t')\right\rangle 
\end{array} \right)\; ,
\end{equation}
obey the linear differential equation given in the
Eq.~(\ref{eq:Bloch}) with ${\bf b}^T= (-\gamma \langle \sigma_z(t')
\rangle, 0, 0)$.

When the counter-propagating pulses are initially in single-photon
Fock-states, the matrix element $\langle 0_a,1_b|\sigma_z(t')|
1\rangle$ where $| 1 \rangle=|0_a,1_b\rangle+|1_a,0_b\rangle$ is
needed. For symmetry reasons, the required element is just
the half of $\langle 1|\sigma_z(t')| 1\rangle$, which can be obtained from the
solution of (\ref{eq:Bloch}) with
\begin{equation}
{\bf s}(t)= \left( \begin{array}{c}
 \langle 1 |\hat\sigma_z| 1 \rangle\\
 \Re{e\left\{\langle 0 |\hat\sigma'_- |1\rangle \right\} }\\
 \Im{m\left\{\langle 0 |\hat\sigma'_- |1\rangle \right\} } 
 \end{array} \right)\; ,\quad 
{\bf b} = \left( \begin{array}{c} 
-\gamma \\ 
2 g(t) \\ 
0 \end{array} \right)\; ,\quad
{\bf B}= \left( \begin{array}{ccc}
-2 \gamma & - 4 g(t) & 0 \\
0 & - \gamma & \omega_0-\omega_A  \\ 
0 & \omega_A-\omega_0 & -\gamma \end{array} \right)\; .
\end{equation}
Note that non-diagonal matrix elements are involved in this set of
equations. One can easily see that closed sets could be derived with
increasing number of equations for higher photon number states, i.e.,
the presented approach is suitable to describe many other initial
states as well. Finally, one gets for the necessary two-time average 
\begin{multline}
 \label{eq:QST_fok}
{\bf s} (t) = \left( \begin{array}{c}
 \langle 1 |\hat\sigma_z(t)  \hat\sigma_z(t')| 1\rangle\\
 \langle 0| \hat\sigma_-'(t) \hat\sigma_z(t')| 1\rangle\\
 \langle 1| \hat\sigma_+'(t) \hat\sigma_z(t')| 0\rangle 
\end{array} \right)\; ,\quad 
{\bf b} = \left( \begin{array}{c} 
-\gamma \langle 1 | \sigma_z(t') | 1 \rangle\\ 
g(t) \\ 
g(t) \end{array} \right)\; ,\\
{\bf B}= \left( \begin{array}{ccc}
-2 \gamma & - 2 g(t) &  -2g(t) \\
0 & i(\omega_0-\omega_A)- \gamma & 0  \\ 
0 & 0  & i(\omega_A-\omega_0)-\gamma \end{array} \right)\; .
\end{multline}

\end{document}